\documentclass[twocolumn,prl,amsmath,aps,groupedaddress,superscriptaddress,
floatfix,showpacs]{revtex4}
\usepackage{amsmath}

\newcommand{\ket}[1]{|#1\rangle}

\begin{document}
\title{Generation of maximally entangled charge-qubit arrays via a cavity mode}
\author{Alexandre Guillaume}
\email{Alexandre.Guillaume@jpl.nasa.gov}
\affiliation{Quantum Computing Technologies Group, Jet Propulsion Laboratory,
California Institute of Technology, Mail Stop 302-306, 4800 Oak Grove Drive,
Pasadena, CA 91109-8099}
\author{Jonathan P. Dowling}
\affiliation{Hearne Institute for Theoretical Physics, Department of Physics
and Astronomy, Louisiana State University,202 Nicholson Hall, Tower Drive
Baton Rouge, LA 70803-4001}
\affiliation{Inst.~for Quantum Studies \& Dept. of Physics, Texas A\&M
University, College Station, Texas 77843-4242}
\date{\today}
\begin{abstract}
We describe an assembly of \textit{N} Cooper-pair boxes (CPB) contained in a
single mode cavity. In the dispersive regime, the correlation between the
cavity field and each Cooper-pair box results in an effective interaction
between CPBs that can be used to dynamically generate maximally entangled
states. With only collective manipulations, we show how to create maximally
entangled quantum states and how to use these states to reach the Heisenberg
limit in the determination of a spectroscopy frequency. This scheme can be
applied to other types of superconducting qubits.
\end{abstract}
\maketitle

The description of the interaction between atoms and quantized modes of the
electromagnetic field in a cavity is called cavity quantum electrodynamics
(cQED). The first experimental studies used flying Rydberg atoms in an rf
resonator. With the advent of quantum computing several other implementations
were developed to mimic the quantum properties of atoms. Among those,
solid-state implementations are especially interesting since they offer several
advantages over real atoms: these artificial atoms properties can be tailored
and their number and location is fixed. We concentrate our discussion on a
superconducting circuit called a Cooper-pair box (CPB). Several experiments
have shown that these circuits behaved like quantum two-level systems
\cite{nakamura1999,vion2002,lehnert2003,duty2004,guillaume2004}.
An architecture has been proposed to realize an on-chip cQED experiment
using a CPB as the artificial atom strongly coupled
to a one-dimensional cavity \cite{blais2004}. This architecture was
implemented and the strength of the coupling was shown to
be indeed stronger than the the different decays constants so that the vacuum Rabi splitting was observed \cite{wallraff2004}. Subsequently the ac-Stark
shift  was measured using a quantum non-demolition technique (QND)
\cite{schuster2004} and the decoherence time $T_2$ evaluated from Ramsey
fringe experiments \cite{wallraff2005}.

The possibility of entangling two Cooper-pair boxes in a cavity QED scheme
was described in reference \cite{blais2004}. Beside its fundamental interest
and applications to quantum information processing, entanglement offers the
additional advantage of allowing improved sensitivity in a quantum-limited
measurement. Interferences between two different polarizations (modes) of
three and four photon entangled states have been observed \cite{mitchell2004,walther2004}. Spectroscopy performed on an assembly
of three beryllium ions also demonstrated a similar improvement in frequency
estimation \cite{leibfried2004}. Using entanglement in solid-state implementations to enable such Heisenberg limited measurements could revolutionize sensor technology with, for instance, electrometers and
magnetometers.

We propose to use the photon-qubit interaction to create an
effective interaction between all the CPBs. We show how this interaction can
be used to generate maximally entangled states and in turn how to use these
states to beat the standard quantum limit when measuring the Ramsey frequency.

The Hamiltonian of a single Cooper-pair box can be expressed as
\cite{makhlin2001a}:
\begin{equation}
	\label{eq:hTLS}
	H_{Q}= -B_z\frac{\sigma_z}{2}-B_x\frac{\sigma_x}{2}
\end{equation}
The $\sigma_k$ are the usual Pauli matrices ($k=x,y,z$) expressed in the charge
basis $\left\{\ket{0},\ket{1}\right\}$. In the linear approximation, the
diagonal term, or bias, depends linearly on the gate charge $n_g$,
$B_z(n_g)=8E_c (1/2-n_g)$ where $E_c$ is the charging energy. So adding an
increment $\delta \hat n_g$ to the gate charge $n_g$ results in a new
Hamiltonian $H_{Q}(n_g+\delta \hat n_g)=H_{Q}(n_g)-( 8E_c\delta \hat n_g/2)
\sigma_z$. We assume this approximation to be valid in a limited range
$0\leq n_g \leq1$ (in units of 2e). The tunneling matrix element $B_x$ is the
Josephson coupling $E_J$. In reference to the case where this quantity is zero
(and therefore the Hamiltonian	diagonal), the point $n_g=1/2$ is called the
\emph{degeneracy point}. The linear approximation is best justified in the
vicinity of this point. For charge qubits, $E_c>E_J$, so that the
Hamiltonian is approximately diagonal in the charge basis at $n_g=0$ and
$n_g=1$. In the remainder of this article, we work at the degeneracy point.

We now consider the case of a Cooper-pair box placed in a single mode
cavity and subject to a quantized field $\delta \hat n_g=n_{c} (a^{\dagger}+a)$
on its gate. We are interested in the strong coupling limit, $g>\kappa$, and
therefore we neglect the cavity decay rate $\kappa$ (and the decay of the qubits
excited state). We express the Hamiltonian in the eigenbasis of
(\ref{eq:hTLS}) and use the rotating wave approximation
(RWA):
\begin{eqnarray}
	\label{eq:hN}
	H=\omega_c (a^{\dagger}a+1/2)+E_J \frac{\sigma_z}{2}+
	\frac{g}{2}(a^{\dagger}\sigma_-+a\sigma_+)	
\end{eqnarray}
Here, we define the coupling constant $g\equiv-4E_c n_{c}$. We can use the
above Hamiltonian to describe several Cooper-pair boxes present in the cavity
when they are not directly coupled. Assuming each CPB experiences the same
coupling to the cavity field, we can rewrite:
\begin{equation}
	\label{eq:hjc}
	H_{N}= \omega_c (a^{\dagger}a+1/2)+E_J S_{z}+
	g (a^{\dagger} S_{-} + a S_{+} )
\end{equation}
We used the notation $S_z=\frac{1}{2}\sum_{i=1}^N \sigma_{z_i}$, $N$ being the number of qubits ($N>1$). Even though the above Hamiltonian is in principle
valid for any $N$, we work far away from the thermodynamic limit $N\rightarrow
\infty$, and assume $N<\omega_c E_J/g^2$. Equation (\ref{eq:hjc}) is also known
as the \textit{N}-atom Jaynes-Cummings Hamiltonian. We use the eigenstates of
$S_z$ to label the total quantum states $\ket{M}_z$ (Dicke states $\ket{J,M}_z$
with $J=N/2$).

In the case of a single atom, the exact diagonalization can be performed
in the so-called dressed-states basis \cite{haroche1992} and leads to the
energy spectrum $E_{\overline{\pm,n}}=(n+1)\omega_c \pm \frac{1}{2}
\sqrt{4g^2(n+1)+\Delta^2}$. The detuning $\Delta$ is defined as 
$E_J-\omega_c$. In the large detuning regime $g/\Delta \ll 1$ the $N$-atom
Jaynes-Cumming Hamiltonian (\ref{eq:hjc}) can be approximately diagonalized
with the transformation :
\begin{equation}
	\label{eq:Ux12}
	U=\exp(g/\Delta (a S_{+}-a^{\dagger}S_{-}))
\end{equation}
We calculate $H'_N= U H_{N} U^{\dagger}$ and get
\begin{gather}
	H'_N \approx 
	\bigl[ \omega_c +\frac{ g^2 }{\Delta} 2S_{z} \bigr]
	a^{\dagger}a +
	\frac{1}{2} \bigl[ E_J +\frac{ g^2 }{\Delta} \bigr]
	2S_{z}
	\nonumber \\
	\label{eq:hNqapprox}
	+\frac{ g^2 }{\Delta}
	(S^2-S_z^2)
\end{gather}
We define $\chi\equiv g^2/\Delta$.
The shift in the resonator frequency (linear in photon number $\hat{n}=
a^{\dagger}a$) is known as the ac-Stark shift. In the second term, the
shift in the atomic frequency is known as the Lamb shift. The last term
shows that a correlation $\chi S_z^2$ exists between qubits when
$N>1$.

Hamiltonian (\ref{eq:hNqapprox}) contains
the nonlinear interaction $H_{sz}=\chi \, S_{z}^2$. It is well known that
this so-called one-axis twisting term can produces a spin-squeezed state when
applied to a coherent spin state \cite{ueda1993}. A \emph{coherent spin state}
is an eigenstate of the angular momentum operator $S_{n}=\mathbf{n} \cdot
\mathbf{S}$ with an eigenvalue $N/2$ [$\mathbf{n}$ is a unit vector pointing
in the direction ($\theta,\phi$)]. The time evolution of such an interaction
term applied to an initial state perpendicular to the $z$ axis, for a time
$t_{sz}=\pi/2\chi$, produces a maximally entangled state (assuming \textit{N}
even. If \textit{N} is odd, another rotation should be added in the process
\cite{molmer1999a}):
\begin{equation}
		\ket{\Psi_1}=\exp(-iH_{sz}\,t_{sz}) \ket{g}=\frac{1}{\sqrt{2}}
		\Big( \ket{g}+ i^{1+N}\ket{e}\Big)
	\end{equation}
we used $\ket{g}\equiv e^{-i S_x \pi/2}\ket{-N/2}_z$ and
$\ket{e}\equiv e^{-i S_x \pi}\ket{g}$. Thus, Hamiltonian (\ref{eq:hNqapprox})
can be used to dynamically produce maximally entangled states.

Provided the rotation can be made in a time substantially smaller than
the entangling time, we propose to start from the ground state $\ket{-N/2}_z$,
perform a $\pi/2$ rotation, say around the axis $x$, then let the
system evolve freely during $t_{sz}=\pi/2\chi$ and finally perform
another $\pi/2$ rotation:
\begin{equation}
		U_N\equiv e^{+i S_x \pi/2} \,	U_{sz}(t_{sz}) \, e^{-i S_x \pi/2}
\end{equation}
$U_N$ can be seen as a generalized $\pi/2$ Ramsey pulse. Starting from an
initial state $\ket{-N/2}_z$, it produces a superposition state $\frac{1}
{\sqrt{2}}(\ket{-N/2}_z+\ket{+N/2}_z)$.
A method has been proposed recently \cite{leibfried2004} in the context
of ion traps, where \emph{one} collective measurement demonstrates the use of
maximally entangled states to reach Heisenberg-limited spectroscopy.
It generalizes the idea of the Ramsey fringe experiment to \textit{N}
particles. First a $\pi/2$ pulse, $U_N$, is applied to the system. Then the
system acquires a phase during a free evolution period that lasts until another
$\pi/2$ Ramsey pulse is applied:
\begin{gather}
	\label{eq:phifinal}
	\ket{\Psi_{final}}=U_N \, e^{-i \, S_z \varphi} \, U_N \ket{-N/2}_z
	\\
	=
	-i \sin(\frac{N \varphi}{2}) \ket{-N/2}_z
	+i^{N+1} \cos(\frac{N \varphi}{2}) \ket{+N/2}_z
\end{gather}
The probability for the system to collapse in the state $\ket{+N/2}_z=
\ket{\uparrow}$ is then simply $P_{\uparrow}=\frac{1}{2}(1+\cos(N\varphi))$.
The phase $\varphi$ can be tuned by controlling the free evolution time
$T$ in between the entangling sequences $U_N$. Alternatively, the
phase can be controlled  by adjusting an offset phase in the $\pi/2$ pulses of
the last $U_N$ sequence \cite{leibfried2004}. From the expression of
$P_{\uparrow}$, we see that it is possible to evaluate the phase with an
uncertainty, at best, equal to $1/N$. This so-called Heisenberg limit
represents an improvement of $1/\sqrt{N}$ over the standard quantum limit
(obtained for a assembly of \textit{N} atoms randomly correlated). This method
has been proposed to improve atomic frequency measurements \cite{bollinger1996}.
For Cooper-pair boxes, the Josephson energy is analogous to the atomic
frequency. So the method proposed here uses entanglement to improve the
accuracy on the measurement of $E_J$ with spectroscopy.

The vast majority of experiments performed with a CPB uses a SQUID
geometry so that the Josephson energy $E_{J_{max}} \cos(\pi \Phi/\Phi_0)$
can be tuned with an external magnetic flux $\Phi$  ($\Phi_0$ is the
flux quantum). Hence, an improvement in Josephson energy  determination
results in an improved flux measurement (if the energy $E_{J_{max}}$ independently measured).
Since Cooper-pair boxes are charge qubits, these devices are not effective
magnetic flux sensors (electric charge and magnetic flux are two quantum
conjugate variables). Therefore, we emphasize that the improvement in
the flux determination is only relative (to the case where CPBs wouldn't
be correlated).

Instead of the fluorescence method usually used in ions traps to read
the probability $P_{\uparrow}$ (or $P_{\downarrow}$), the QND measurement
proposed and used in reference \cite{blais2004,wallraff2004,schuster2004,
wallraff2005} is more appropriate to our scheme.
Beside the term $\chi S_z^2$, equation (\ref{eq:hNqapprox}) contains
another term, $2\chi S_z a^{\dagger}a$, describing the interaction
field-qubits. Because $S_z$ commutes with this interaction and the Hamiltonian
(\ref{eq:hNqapprox}), a quantum non-demolition measurement can be performed
and $S_z$ measured.
If a continuous wave signal at the resonator frequency $\omega_c$ is applied
on the system, the term corresponding to the ac-Stark effect will introduce
an equal but opposite phase shift for the two components $\ket{\pm N/2}_z$.
If we start from an initial state
$\big( \alpha \ket{+N/2}_z+\beta  \ket{-N/2}_z \big) \ket{\theta}$
where $\ket{\theta}$ is a coherent state, the time evolution corresponding
to the interaction $(g^2/\Delta) \, a^{\dagger}a \, \sigma_z$ will create a
state $\alpha \ket{\theta, +N/2}_z+\beta  \ket{-\theta, -N/2}_z$ where
$\tan\theta=2g^2/\kappa \Delta$ \cite{blais2004} ($\kappa$ is the cavity decay
rate $\omega_c/Q$).
The phase difference between the two states $\ket{\pm N/2}_z$ can be
measured by a transmission measurement of the photons through the cavity.
The probability $P_{\uparrow}$ is extracted from the
time dependence of $\theta$. This measurement can be either continuous
(weak) or pulsed (stronger). The scheme proposed here is scalable to any
N (as long as the qubits are only coupled together though the cavity mode).

For a 24 mm long cavity \cite{wallraff2004}, we can reasonably assume that an
assembly of four CPBs spaced by 100 $\mu$m from each other, will experience a
homogeneous field at the central antinode and at the same time that the CPBs
will not be directly coupled. For a minimum detuning
of approximately 100 MHz and a coupling strength of 17 MHz, the time needed to
generate a maximally entangled state would be around 346 ns. The squeezing time
obviously has to be smaller than the decoherence time. The time $T_2$ has been
evaluated to be 500 ns in the case of one qubit \cite{wallraff2005}. The
squeezing time could be decreased further by increasing the couplings strength
so that the condition of large detuning $g\ll \Delta$ for which eq.
(\ref{eq:hNqapprox}) is valid would still hold. In order to minimize the effect
of the measurement, the probability $P_{\uparrow}$ can be read after the last
$\pi/2$ pulse \cite{wallraff2005}.
We conclude that with existing technology it is possible to create maximally
entangled states between N (four) charge qubits in a superconducting cavity
QED and improve by a factor $\sqrt{N}$ ($\sqrt{2}$) the accuracy on the
measurement of the Ramsey frequency. 

The method we proposed can be simplified further. The general sequence
(\ref{eq:phifinal}) requires four $\pi/2$ pulses (two per general Ramsey
pulse $U_N$). This number can be decreased by half. For this purpose,
we examine the situation of atomic type spectroscopy \cite{haroche1992}
performed on the  system at the degeneracy point. The effect of a classical
periodic drive on the gate can be described by adding
$\varepsilon (e^{-i\omega t} S_+ + e^{i\omega t} S_-)$ to equation
(\ref{eq:hjc}). In the rotating frame at $\omega$:
\begin{gather}
	H_{\omega}= \delta_1 S_{z}+\delta_2 \, a^{\dagger}a+
	g (a^{\dagger} S_{-} + a S_{+} )+\nonumber\\
	\label{eq:hjccwrot}
	\varepsilon ( S_+ + S_-)
\end{gather}
with $\delta_1=E_J-\omega$ and $\delta_2=\omega_c-\omega$. When
the drive is resonant with the qubits, $\delta_1=0$, this interaction term can
be expressed in the interaction picture \cite{agarwal2003}:
\begin{gather}
	\label{eq:hintcw}
	H_I= 
	g ( S_{x} + \frac{S_{z}-iS_{y}}{2} \, e^{i\, 2\varepsilon t}
	- \frac{S_{z}+iS_{y}}{2} \, e^{-i\, 2\varepsilon t})
	a^{\dagger}e^{i\,\delta_2 t}
	\nonumber \\
	+ h.c.
\end{gather}
We restrict ourselves to the weak drive limit $\delta_2 > \varepsilon$. Using
a method described by D.F. James \cite{james2000}, we average this interaction
over a time $T\gg \pi/\varepsilon$ and derive the effective Hamiltonian:
\begin{gather}
	H_{\rm eff}=\frac{g^2}{\delta_2} S_{x}^2 
	\nonumber \\
	+\Big(\frac{g}{2}\Big)^2 \frac{1}{\delta_2-2\varepsilon}
	\big[ (S^2 -S_{x}^2)+(1+2a^{\dagger}a)S_{x} \big]
	\nonumber \\
	\label{eq:heff}
	+\Big(\frac{g}{2}\Big)^2 \frac{1}{\delta_2+2\varepsilon}
	\big[ (S^2 -S_{x}^2)-(1+2a^{\dagger}a)S_{x} \big]
\end{gather}
The advantage of this expression over	equation (\ref{eq:hNqapprox}) is that
it is expressed in a basis orthogonal to the original $z$ axis so that the
one axis twisting term is now proportional to $S_x^2$. Starting from
the ground state $\ket{-N/2}_z$ and turning on the interaction $H_{\rm eff}$
for a time  approximately equal to $(\pi \delta_2)/(4 g^2)$ will produce a
maximally entangled state. So, with the same initial state, only one
manipulation is needed to generate a GHZ state with a classical periodic drive whereas two $\pi/2$ pulses are needed with the expression (\ref{eq:hNqapprox}).
The coupling  $g$ should also be smaller than the detuning $\delta_2$ so that
the entangling time is long enough compare to the average time $T$ after which
(\ref{eq:heff}) is valid. This effective Hamiltonian is valid for any system
described by a N-atom Jaynes-Cummings Hamiltonian subject to a weak period
drive resonant with the qubits.
Shifts analogous to one found in the Hamiltonian (\ref{eq:hNqapprox}), in the
atomic and especially the resonator frequency are present in (\ref{eq:heff}).
Thus, the same kind of quantum non-demolition measurement as proposed in \cite{blais2004} can be performed.

In this work, we described a collection of \textit{N} Cooper-pair boxes
contained in a single mode cavity. At the degeneracy point, we showed that the
diagonal Hamiltonian contains an effective qubit-qubit interaction mediated by the cavity mode, of the form $\chi S_z^2$. This interaction can be used to
prepare maximally entangled states. We adapt a method used in ion traps
to demonstrate the use of maximally entangled state to reach the Heisenberg
limit in the Ramsey frequency determination. We estimate that such an
experiment is possible with existing technology.
Additionally, we showed that the $N$-atom Jaynes-Cummning Hamiltonian, with a
weak classical drive on resonance with the qubits, exhibits the same properties
as the same Hamitonian without a drive, but expressed in an  basis orthogonal to the eigenbasis of the system without drive. This approach decreases the
number of pulses needed in our scheme to demonstrate entanglement.
The results presented in this work are valid for two-level systems that
can be described with the Hamiltonian (\ref{eq:hTLS}). For instance, it
can be applied to an assembly of rf-SQUIDs. This scheme is especially
interesting for solid state implementation since it only requires 
collective manipulations.

\begin{acknowledgments}
The research described in this paper was carried out at the Jet Propulsion Laboratory, California Institute of Technology, under a contract with the National Aeronautics and Space Administration.
A.G. would like to acknowledge the National Research Council and NASA
code S. Also J.P.D. would like to acknowledge the Hearne Foundation, the
National Security Agency, the Advanced Research and Development Activity and
the Army Research Office.
\end{acknowledgments}




\begin{thebibliography}{10}

\bibitem{nakamura1999}
Y. Nakamura, Y. Pashkin, and J. Tsai, Nature {\bf 398},  786  (1999).

\bibitem{vion2002}
D. Vion {\it et~al.}, Science {\bf 296},  889  (2002).

\bibitem{lehnert2003}
K. Lehnert {\it et~al.}, Phys. Rev. Let. {\bf 90},  027002  (2003).

\bibitem{duty2004}
T. Duty, D. Gunnarsson, K. Bladh, and P. Delsing, Phys. Rev. B {\bf 69},
  140503(R)  (2004).

\bibitem{guillaume2004}
A. Guillaume {\it et~al.}, Phys. Rev. B {\bf 69},  132504  (2004).

\bibitem{blais2004}
A. Blais {\it et~al.}, Phys. Rev. A {\bf 69},  062320  (2004).

\bibitem{wallraff2004}
A. Wallraff {\it et~al.}, Nature {\bf 431},  162  (2004).

\bibitem{schuster2004}
D.~I. Schuster {\it et~al.}, Phys. Rev. Let. {\bf 94},  123602  (2005).

\bibitem{wallraff2005}
A. Wallraff {\it et~al.}, cond-mat/0502645 (unpublished).

\bibitem{mitchell2004}
M. Mitchell, J. Lundeen, and A. Steinberg, Nature {\bf 429},  161  (2004).

\bibitem{walther2004}
P. Walther {\it et~al.}, Nature {\bf 429},  158  (2004).

\bibitem{leibfried2004}
D. Leibfried {\it et~al.}, Science {\bf 304},  1476  (2004).

\bibitem{makhlin2001a}
Y. Makhlin, G. Sch{\"{o}}n, and A. Shnirman, Rev. Mod. Phys. {\bf 73},  357
  (2001).

\bibitem{haroche1992}
S. Haroche,  in {\em Fundamental Systems in Quantum Optics}, edited by J.
  Dalibard, J. Raimond, and J. Zinn-Justin (Elsevier Science Publishers, New
  York, 1992).

\bibitem{ueda1993}
M. Kitagawa and M. Ueda, Phys. Rev. A {\bf 47},  5138  (1993).

\bibitem{molmer1999a}
K. M{\o}lmer and A. S{\o}rensen, Phys. Rev. Let. {\bf 82},  1835  (1999).

\bibitem{bollinger1996}
J.~J. Bollinger {\it et~al.}, Phys. Rev. A {\bf 54},  R4649  (1996).

\bibitem{agarwal2003}
E. Solano, G.~S. Agarwal, and H. Walther, Phys. Rev. Let. {\bf 90},  027903
  (2003).

\bibitem{james2000}
D. James, Fortschr. Phys. {\bf 48},  823  (2000).

\end{thebibliography}
\end{document}